\documentclass{aa520}
\usepackage[dvips]{graphics}
\def\gtrsim{\lower.45ex\hbox{$\;\buildrel>\over\sim\;$}}
\def\ltrsim{\lower.45ex\hbox{$\;\buildrel<\over\sim\;$}}
\def\micron{\,$\mu$m}
\def\Lp{$L^\prime$}
\def\cii{[C\,{\sc ii}]}
\def\oi{[O\,{\sc i}]}
\def\hii{H\,{\sc ii}}
\def\g0{$G_0$}
\begin{document}
\title{Hot dust in normal star-forming galaxies: $JHK$\Lp\ photometry
of the ISO Key Project sample
\thanks{Based on data obtained at TIRGO, Gornergrat, Switzerland}}
\author{ L. K. Hunt\inst{1}, C. Giovanardi\inst{2}, G. Helou\inst{3}}
\offprints{L.K. Hunt}
\institute{ Istituto di Radioastronomia-Firenze/CNR, Largo E. Fermi 5, 50125 Firenze - Italy
email: hunt@arcetri.astro.it
\and
INAF-Osservatorio Astrofisico di Arcetri,
Largo E. Fermi 5, 50125 Firenze - Italy
email: giova@arcetri.astro.it
\and
IPAC-Caltech,
Pasadena, CA - USA  email: gxh@ipac.caltech.edu }
\date{Received ; accepted }
\titlerunning{Hot dust in star-forming galaxies}
\authorrunning{Hunt, Giovanardi, \& Helou} 
\abstract{We present $JHK$ and 3.8\micron\ (\Lp) photometry of 26 galaxies
in the {\it Infrared Space Observatory} (ISO) Normal Galaxy Key 
Project (KP) sample and of seven normal ellipticals with
the aim of investigating the origin of the 4\micron\ emission.
The majority of the KP galaxies, and all the ellipticals, have
$K-L\,\ltrsim\,1.0$,
consistent with stellar photospheres plus moderate dust extinction.
Ten of the 26 KP galaxies have $K-L\,\gtrsim\,1.0$, 
corresponding to a flat or rising 4\micron\ continuum, 
consistent with significant emission from hot dust at 600--1000\,K.
$K-L$ is anticorrelated with ISO flux 
ratio $F_{6.75}/F_{15}$, weakly correlated with line ratio \oi/\cii,
but not with \cii/FIR or IRAS ratio $F_{60}/F_{100}$.
Photodissociation-region models for these galaxies show that the hot dust 
responsible for red $K-L$ resides in regions of high pressure and intense 
far-ultraviolet radiation field. 
Taken together, these results suggest that star formation in normal star-forming
galaxies can assume two basic forms: 
an ``active'', relatively rare, mode characterized by hot dust, suppressed 
Aromatic Features in Emission (AFEs), high pressure, and intense radiation field;
and the more common ``passive'' mode 
that occurs under more quiescent physical conditions,
{\it with} AFEs, and {\it without} hot dust.
The occurrence of these modes appears to only weakly depend on the
star-formation rate per unit area.
Passive star formation over large scales makes up the bulk of 
star-forming activity locally, while the ``active'' regime may dominate at 
high redshifts.
}
\maketitle
%
%
\section{Introduction}
The {\it Infrared Space Observatory} (ISO) mission has provided an unprecedented
view of the interstellar medium (ISM) in galaxies from
the near-infrared (NIR) continuum between 3 and 5\micron\  to
the C$^+$ fine-structure transition at 158\micron\ and beyond.
Infrared spectra (3 to 12\micron)
obtained by ISO-PHOT reveal the mid-infrared (MIR) emission from {\it normal
galaxies} to be characterized by the Aromatic Features in Emission 
(AFEs) at 3.3, 6.2, 7.7, 8.6, and 11.3\micron, and an
underlying continuum which contributes about half the
luminosity in the 3 to 12\micron\ range (Helou et al. \cite{h2000}).

However, the ISO data have also raised
many questions about the ISM constituents.
In particular,
the nature of the dust responsible for the AFEs is not yet clear,
and the debate over three-dimensional Very Small Grains (VSGs),
two-dimensional Polycyclic Aromatic Hydrocarbon
molecules (PAHs), or Amorphous Hydrogenated Carbon particles
(HACs), and their relative contribution in different environments
is still underway (Jenniskens \& D\'esert  \cite{jenniskens}; Lu \cite{lu};
Cesarsky et al. \cite{cesarsky98}).
Moreover, the relationship between the carriers of the AFEs
and the underlying continuum remains obscure
(Boulanger et al. \cite{boulanger}, Helou et al. \cite{h2000}).

A related question is the relative contributions of dust in the ISM,
stellar photospheres, and circumstellar dust to the MIR radiation in spirals
and ellipticals.
IRAS data already convincingly showed that the MIR emission in spirals is
mainly due to a continuum from small grains transiently 
heated to high temperatures and AFEs.
On the other hand, in elliptical galaxies the ISM-to-stellar ratio is
generally low, so that photospheric
and circumstellar emission from evolved red giants may contribute
significantly to the MIR
(Knapp et al. \cite{knapp}, Mazzei \& de Zotti \cite{mazzei}).
It has been argued, though, that the same IRAS color-color relation
(e.g., Helou \cite{h86}) holds for ellipticals and spirals, which
implies a similar ISM origin (Sauvage \& Thuan \cite{st}).

The continuum component between 3 and 5\micron\ can be extrapolated
fairly well to the continuum level at 9--10\micron\ (Helou et al. \cite{h2000}),
which suggests that some of the 3--5\micron\ continuum must be produced
by the ISM.
Nevertheless, a portion of the 3--5\micron\ continuum must arise
from stellar photospheres since they dominate the emission at
1--2\micron .
Although the signal-to-noise in the 3--5\micron\ ISO spectra is relatively
low, that is the spectral region where the MIR spectra of the 43 galaxies
observed by Helou et al. (\cite{h2000}) show the most significant 
galaxy-to-galaxy variation.
To understand the physical origin of this, and assess the
relationship of the MIR continuum to the ubiquitous AFEs,
it is necessary to separate the ISM contribution from that of
stellar photospheres and circumstellar dust.
To this end,
we have acquired $JHK$\Lp\ photometry of the galaxies studied
in Helou et al. (\cite{h2000}), which 
constitutes one of the ISO Normal Galaxy Key Projects.
A variety of ISO observations have been secured for
this sample, including 
mid-infrared spectra (Helou et al. \cite{h2000}), ionic and atomic fine-structure line
fluxes
(Malhotra et al. \cite{m1997}, Malhotra et al. \cite{m2001}), mid-infrared maps at 7 and 15\micron,
and a small number of 4.5\micron\ images (Dale et al. \cite{d2000}).

The rest of the paper is organized as follows:
Sect. \ref{obs} describes the observations, reduction, and photometric
calibration.
We compare the $JHKL$ colors of the sample galaxies with their ISO 
properties in Sect. \ref{iso},
and with photodissociation-region models in Sect. \ref{pdr}.
Section \ref{nature} discusses the nature of the 4\,\micron\ continuum,
and our interpretation of these results
in terms of ``active'' and ``passive'' star formation.
\begin{figure*}
{\rotatebox{180}{\includegraphics*[100,100][570,820]{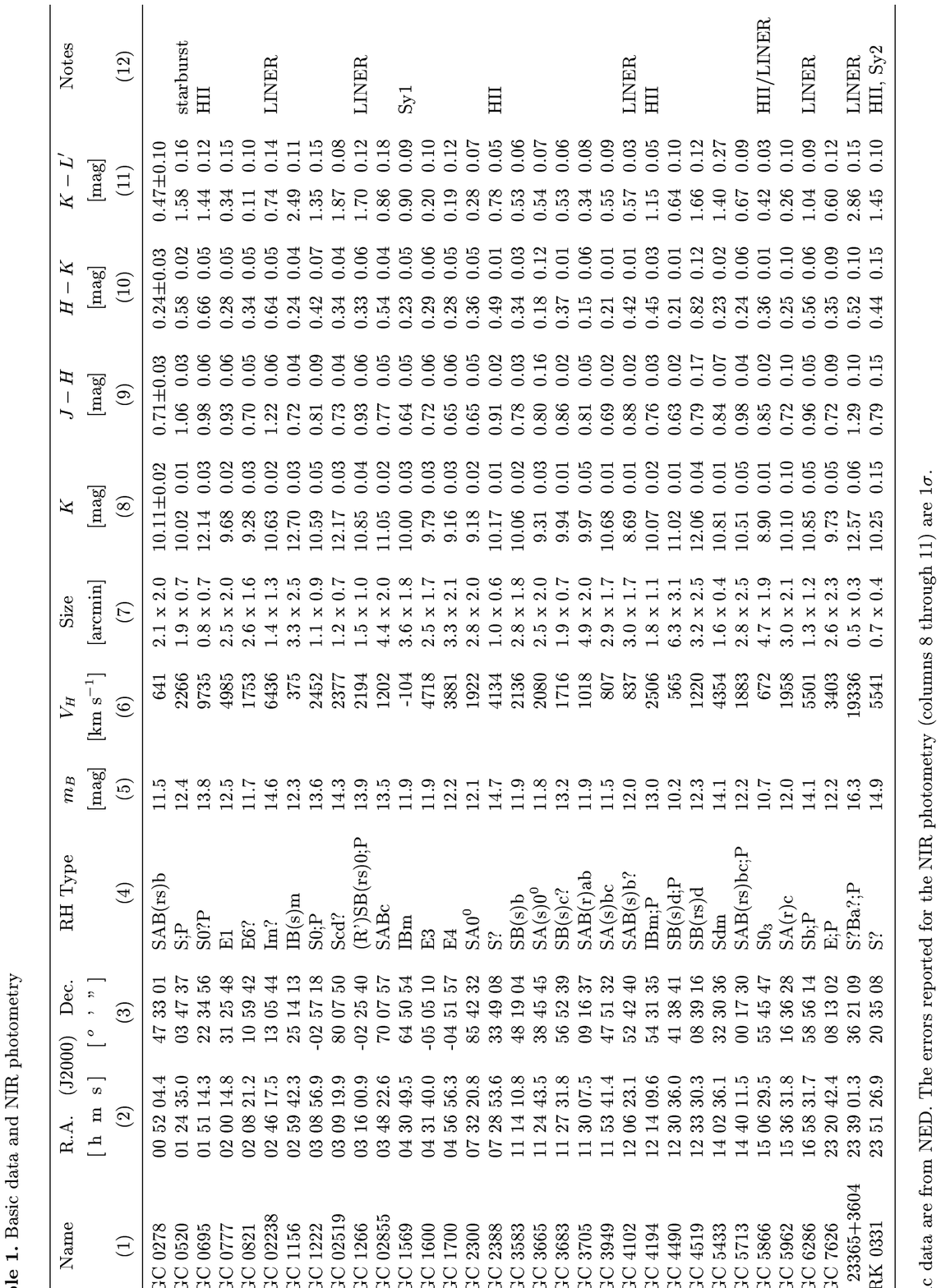}}}
\label{tbl_phot}
\end{figure*}
\section{The sample, the observations, and the colors\label{obs}}
The ISO Key Project for Normal Galaxies (PI: G. Helou, KP) was proposed
to study the ISM of a broad range of normal galaxies using several
instruments aboard ISO.
The sample, described in Malhotra et al. (\cite{m2001}), is designed
to capture the great diversity among galaxies, in terms
of morphology, luminosity, far-infrared-to-blue luminosity ratio $FIR/B$, 
and IRAS colors.
The project obtained ISO observations of 69 galaxies, including
nine relatively nearby and extended objects.
The remaining 60 galaxies cover the full range of observed morphologies,
luminosities, and star-formation rates seen in normal galaxies,
and includes five dwarf irregulars discussed in Hunter et al. (\cite{hunter}). 
Two of these, NGC\,1156 and NGC\,1569, are part of our observed sample,
and are also the two nearest objects (6.4 and 0.9\,Mpc, respectively).
The median distance of the KP sample (including these nearby objects, and 
with $H_0$\,=\,75\,km/s/Mpc) is 34\,Mpc (Dale et al. \cite{d2000}).
This means that with the 14\arcsec\ observing aperture we are sampling
a central region roughly 1 kpc in radius.
We observed 26 KP galaxies.

We also selected seven 
quiescent ellipticals from the sample of early-type galaxies observed with ISO
by Malhotra et al. (\cite{m2000}),
or with published $JHK$ photometry (Frogel et al. \cite{frogel}).
Several of the latter do not have the rich array of ISO
observations obtained for the KP galaxies.

\subsection{Observations}
The observations were acquired at the 1.5\,m f/20 Infrared Telescope
at Gornergrat (TIRGO\footnote{The Infrared Telescope at Gornergrat (Switzerland) is
operated by CAISMI-CNR, Arcetri, Firenze.}), with a single-element InSb detector.
The photometer is equipped with
standard broadband filters ($J$ 1.2\micron, $H$ 1.6\micron, $K$ 2.2\micron,
and \Lp\ 3.8\micron) with diaphragms in the focal plane defining
the aperture dimensions.
The galaxy coordinates were taken from NED\footnote{The NASA/IPAC 
Extragalactic Database (NED)
is operated by the Jet Propulsion Laboratory,
California Institute of Technology, under contract with the U.S.
National Aeronautics and Space Administration.} and checked with
the DSS\footnote{The DSS was produced at
the Space Telescope Science Institute under U.S. Government grant
NAG W-2166.}; all observations were acquired with a 14\arcsec\ aperture,
after maximizing the infrared signal.
Sky subtraction was performed with a wobbling secondary at frequencies
that ranged from 2.1 to 12.5\,Hz, according to the integration time for
the individual measurement.
The modulation direction was in a EW direction, with an amplitude of
roughly 3\arcmin.
Beam switching was used to eliminate linear variations in 
sky emission.

Photometric calibration was achieved by similarly observing several 
standard stars nightly from the CIT (Elias et al. \cite{elias}) and the ARNICA 
(Hunt et al. \cite{hunt98}) standard lists.
Nightly scatter of the photometric zero point was typically 5\% or better
in $JHK$ and 8\% or better in \Lp.

The final photometry for the 33 galaxies observed
is reported, together with their basic data, in Table \ref{tbl_phot}. 
In what follows, we have transformed $K-$\Lp\ to $K-L$ (this last
is centered at 3.5\micron) using the transformation by Bessell \& Brett (\cite{bessell}),
and we will use $K-L$ to denote such colors. In the various plots
and when testing for correlations, the NIR data have also been corrected
for {\it i)} Galactic extinction according to Schlegel et al. (\cite{schlegel}) and 
Cardelli et al. (\cite{cardelli}), and {\it ii)} K dimming using the precepts mentioned
in Hunt \& Giovanardi (\cite{hg92}). 

\begin{figure}
\resizebox{\hsize}{!}{\rotatebox{0}{\includegraphics*[10,140][510,700]{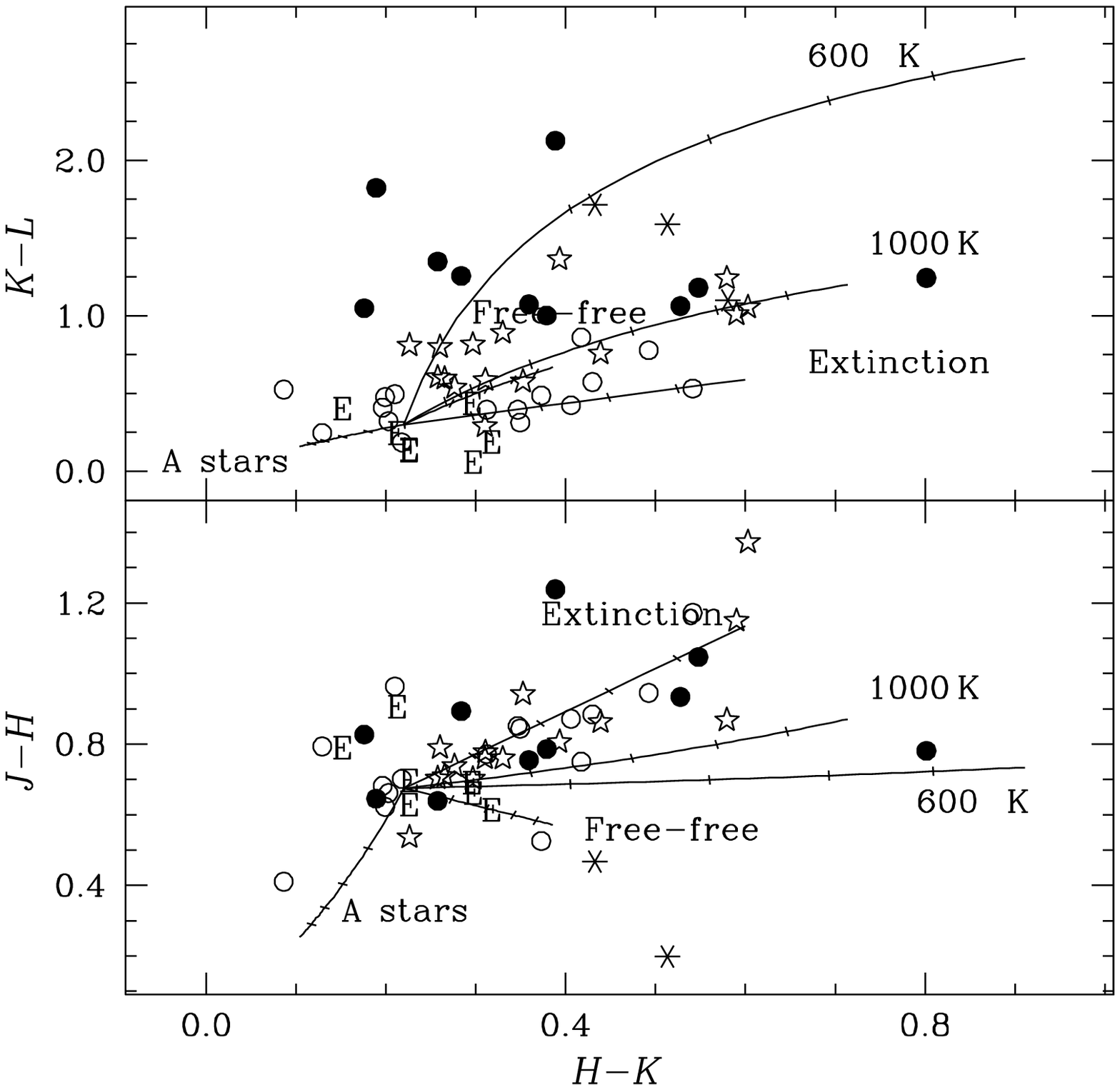}}}
\caption{NIR color-color diagram; the 
$K-L$ color in the top panel is transformed from $K-L'$ 
as described in the text.
Filled symbols refer to galaxies with $K-L > 1.0$.
The 7 quiescent ellipticals are indicated with an E. 
Shown as stars are \hii\ galaxies taken from Glass \& Moorwood (\cite
{glass})
and starbursts from Hunt \& Giovanardi (\cite{hg92}), and shown as asterisks the BCDs 
(see text). Mixing curves show how the colors change when various 
physical processes increasingly contribute to the emission 
observed (see Hunt \& Giovanardi \cite{hg92}); the end points of
the mixing curves indicate equal $K$-band contributions from stars and the
process (hot dust, ionized gas, A stars).
The tick marks on the extinction line show unit $A_V$ magnitude increments.
\label{fig:jhkl} }
\end{figure}
\subsection{$JHKL$ colors \label{colors}}
The NIR colors of the observed galaxies are shown in Fig. \ref{fig:jhkl}. 
The filled symbols represent those (10) galaxies 
with $K-L$\,\gtrsim\,1.0, that is a rising continuum (see below).
Also shown in the diagram are \hii\ galaxies taken from Glass \& Moorwood (\cite{glass})
and starbursts from Hunt \& Giovanardi (\cite{hg92}).
Three low-metallicity blue compact dwarfs (BCDs) are also shown as asterisks:
NGC\,5253 (Glass \& Moorwood \cite{glass}),
II\,Zw\,40 (Thuan \cite{thuan83}), but no $J$), and
SBS\,0335-052 (Hunt et al. \cite{hunt2001}).
The BCDs have the most extreme colors since they tend to be quite blue
in $J-H$ (because of low metallicity and youth), and red in $K-L$ (because of 
ionized gas and hot dust).

Mixing curves show how the colors change when various physical processes
increasingly contribute to the emission observed (see Hunt \& Giovanardi \cite{hg92});
they illustrate that, on the basis of NIR color, it is possible to distinguish 
among stellar photospheric emission,
``passive'' dust in extinction, and ``active'' dust in emission.
$H-K$ tends to be red ($ > 0.35$) for both dust extinction and dust
emission, while $K-L$ is red only because of emission by dust or ionized gas 
(see the free-free line).
Thus red $K-L$ can be used to signal a substantial 
contribution from hot dust emission in the observing aperture.
The 600 and 1000\,K mixing curves span the observed colors quite well,
and although the VSG emission is not thermalized, the curves show that 
the NIR colors are consistent with dust at these temperatures.

The $K-L$ color can also provide an estimate of the slope of the
spectral energy distribution (SED) between 2 and 4\micron.
Using the zero-magnitude fluxes given in Koornneef (\cite{koornneef}), and
assuming a power-law dependence $f_\nu \propto \nu^\alpha$, we find that
$K-L\,=\,-0.50\,\alpha\ +\ 0.95$.
Therefore\footnote{The
intercept climbs from 0.95 to 1.10 if we adopt instead the scale of
Wilson et al. (\cite{wilson}).}, 
$K-L\,\gtrsim\,1.0$ is where the
slope $\alpha$ changes sign and becomes negative, signifying a 
continuum $f_\nu$ rising with increasing wavelength; 
when $K-L\,<\,1.0$, $\alpha$ is positive. 
The mean $\alpha$ averaged over the entire KP sample is $>$\,0,
with $f_\nu\,\propto\,\nu^{+0.65}$ (Helou et al. \cite{h2000}). 
If we include the ellipticals, the median $K-L$ of our 33 galaxies is
0.50, which gives $\alpha\,=\,0.90$, steeper than, but consistent with,
Helou et al. (\cite{h2000}).
If only the ISO Key Project galaxies are considered,
median $K-L\,=\,0.68$, corresponding to $\alpha\,=\,0.54$, remarkably close to
that reported in Helou et al.
The median $K-L$ for the ellipticals only is 0.21, which corresponds to
a very steep falling continuum with $\alpha\,=\,1.5$.

The mixing curves shown in Fig. \ref{fig:jhkl}
illustrate what fraction of the observed 
flux is due to hot dust; they assume a mix of stellar photospheres 
with intrinsic stellar color $(K-L)_*\,=\,0.3$,
or $(K-L^\prime)_*\,=\,0.5$,
plus dust emission.
When the SED is flat at 4\,\micron \,($K-L\,\approx\,1.0$, slope $\alpha\,=\,0$), 
600\,K hot dust comprises roughly 5\% of the total $K$-band flux;
hotter dust (e.g., 1000\,K) would constitute 30\%.
It is unlikely that flat or rising continua are due to free-free emission
from ionized gas, since even with a 50\% emission fraction from gas, 
the continuum is still
falling [$(K-L)\,=\,0.7$].
This means that $K-L$, or alternatively the slope of the 4\micron\ continuum, 
is a remarkably sensitive diagnostic of hot dust: small fractions (5--30\%) of 
dust emission cause large variations (0.6\,mag) in the $K-L$ color.
A 50/50 $K$-band mix of hot dust and stars would produce a $K-L$ color
between 1.2 and 2.7, depending on the dust temperature;
NGC\,4519, the galaxy with the reddest $H-K$ (=\,0.82), may contain such a mix.

\section{Comparison with ISO photometry and spectra \label{iso}}

If the non-stellar 3.8\micron\ flux is truly associated with hot dust,
we would expect to find correlations between ISO MIR observations and \Lp,
if both arise from a similar small-grain population.
Also, observations of the two most important photo-dissociation region
(PDR) cooling lines, \cii\ and \oi, suggest that
gas and dust temperatures increase together
(Malhotra et al. \cite{m2001});
we might therefore expect trends with these line fluxes and ratios and $K-L$.

\subsection{Broadband fluxes \label{isoflux}}

We first checked that our ground-based (small-aperture) \Lp\ fluxes
were broadly consistent with the ISO fluxes at
$F_{6.75}$ and $F_{15}$ (Dale et al. \cite{d2000}). 
All three are mutually correlated (not shown), 
except for two cases (NGC\,278 and NGC\,5713)
where the ISO flux is substantially larger than expected from the general
trend.
This discrepancy is almost certainly due to
aperture effects,
since the ground-based data are acquired in a 14\arcsec\ aperture,
while the ISO values published in Dale et al. (\cite{d2000}) are total
fluxes as extrapolated from curves of growth.
NGC\,1569 is generally an outlier; however, its properties are
similar to those of other low-metallicity dwarf irregulars
(see Hunter et al. \cite{hunter}).
The other dwarf irregular in our observed sample is NGC 1156,
again a clear outlier in most of the subsequent plots and correlations.
As noted in Sec. \ref{obs}, these galaxies are also the two closest
galaxies, which considerably complicates the comparison of the ISO
and our ground-based 14\arcsec\ aperture data.

\begin{figure}
\hspace{-0.5cm}
\resizebox{\hsize}{!}{\rotatebox{0}{\includegraphics*[30,150][570,700]{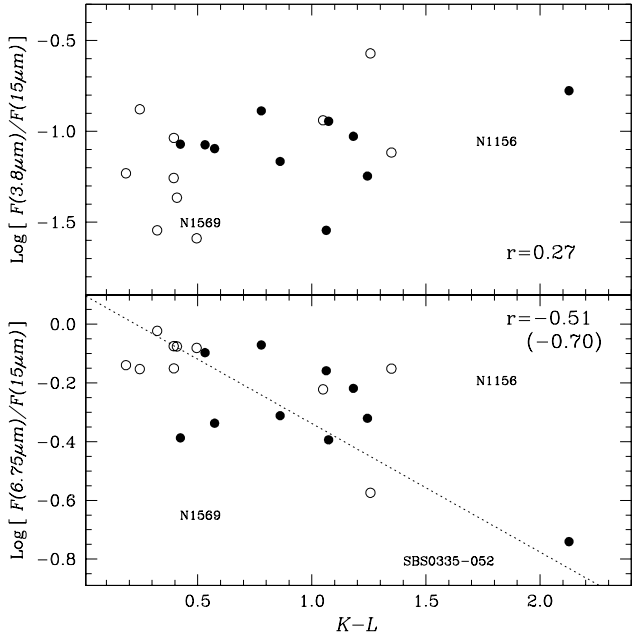}}}
\caption{Hybrid ground-based/ISO color-color plot: the top panel shows
the ratio of the 3.8\micron\ flux (not transformed to $L$) and $F_{15}$
plotted against $K-L$,
and the bottom panel the ratio of $F_{6.75}$ and $F_{15}$ plotted
against $K-L$.
The dotted line shows the best-fit regression; the correlation coefficients
are shown in each panel (parentheses denote the coefficient
without the two nearest galaxies NGC\,1569 and NGC\,1156). 
Solid dots are used when the corrected $H-K>0.35$.
\label{fig:gen} }
\end{figure}

We next compare ISO colors with $K-L$ in Fig. \ref{fig:gen}
where $F_{6.75}/F_{15}$ and the ``hybrid'' color 
$F_{3.8}/F_{15}$ are plotted against $K-L$. 
Ground-based $K-L$ turns out to be correlated (2.5\,$\sigma$)
with $F_{6.75}/F_{15}$\footnote{If the nearest objects, NGC\,1569 and NGC\,1156, 
are eliminated from the regression the significance becomes 4.0\,$\sigma$.
If, in addition, the reddest $K-L$ value (IRAS\,23365$+$3604)
is eliminated, the significance becomes 2.1\,$\sigma$.}, 
but is uncorrelated with the hybrid color.
The sense of the $F_{6.75}/F_{15}$ vs. $K-L$ correlation is
such that {\it redder} $K-L$ implies {\it lower} $F_{6.75}/F_{15}$.
NGC\,1569 stands out since it has a very low $F_{6.75}/F_{15}$
ratio for its $K-L$ color. 
As an extreme case,
in Fig. \ref{fig:gen}, we have also plotted SBS\,0335-052
(not considered in the correlation), 
using the ISO data from Thuan et al. (\cite{thuan99}) and $K-L$ from 
Hunt et al. (\cite{hunt2001}).
SBS\,0335-052 is an unusual BCD with 1/40 solar metallicity, 
and the ISO spectrum shows no AFEs; not surprisingly 
Fig. \ref{fig:gen} shows that
this galaxy has a lower $F_{6.75}/F_{15}$ ratio than any of the
galaxies in the ISO sample.

\begin{figure}
\resizebox{\hsize}{!}{\rotatebox{0}{\includegraphics*[40,150][570,700]{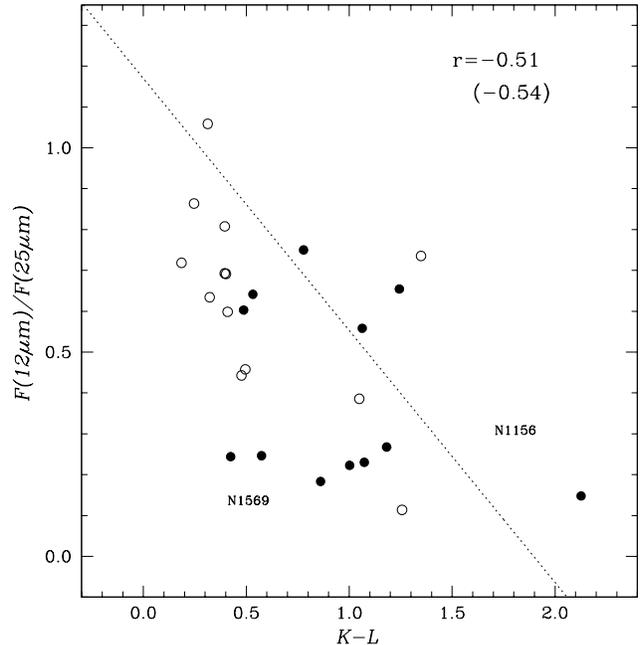}}}
\caption{IRAS 12$\mu m$/25$\mu m$ flux ratio vs. $K-L$.
No IRAS upper limits are shown. Solid points refer to galaxies with a corrected
$H-K>0.35$. 
The best-fit regression is shown together with the correlation coefficient
(in parentheses is that without NGC\,1156 and NGC\,1569, the two nearest galaxies).
\label{fig:kl1225} }
\end{figure}
Since dust temperatures in large ``classical'' grains are connected with
the IRAS flux ratio $F_{60}/F_{100}$, and since this ratio and 
$F_{6.75}/F_{15}$ are anticorrelated (Dale et al. \cite{d2000}), we 
might expect $K-L$ to also be anticorrelated with $F_{60}/F_{100}$. 
We found no such correlation but note, instead, that 
IRAS $F_{12}/F_{25}$ is anticorrelated 
(2.6\,$\sigma$) with $K-L$\footnote{If the 
object with the reddest $K-L$ (IRAS\,23365$+$3604)
is eliminated from the regression, the significance becomes 2.4\,$\sigma$.}, 
as shown in Fig. 
\ref{fig:kl1225}\footnote{Although these galaxies were selected 
to be sufficiently distant,
given the size of the IRAS beams compared to our ground-based photometry,
aperture effects almost certainly contribute significantly to the scatter.}.
Together with the trend with $K-L$ and $F_{6.75}/F_{15}$, this 
means that the presence of hot dust is:
{\it a)}~usually linked to the suppression of AFEs which dominate the 
$12\,\mu$m band (Helou et al. \cite{h91}), 
and {\it b)}~largely independent of the temperature and 
characteristics of the large grains.

\subsection{Line measurements \label{isoline}}

In PDRs and \hii\ regions, gas and dust are intimately related.
In ionized regions, gas and dust compete for far-ultraviolet (FUV) photons, but
the neutral gas in PDRs is heated predominantly by photoelectrons from
small dust grains (Watson \cite{watson}; Hollenbach \& Tielens \cite{hb97}).
Neutral gas is cooled primarily by atomic and ionic fine-structure
lines, \cii\ (158\micron) and \oi\  (63\micron), and these lines
can be used as diagnostics for the physical conditions in the PDR
gas (Tielens \& Hollenbach \cite{th85}):
\begin{figure}
\resizebox{\hsize}{!}{\rotatebox{0}{\includegraphics*[30,150][570,680]{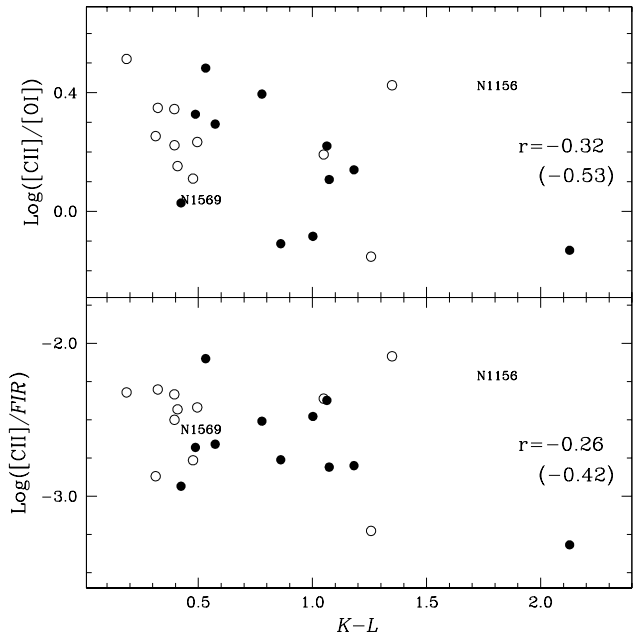}}}
\caption{ISO emission-line ratios vs. $K-L$. The top panel shows
the log of the ratio of \cii\,(158\,$\mu$m)
and \oi\,(63\,$\mu$m) line fluxes vs. $K-L$,
and the bottom panel the log of the ratio of \cii\ and FIR vs.  $K-L$.
The correlation coefficients are shown in each panel
(the correlation coefficient in parentheses is that without NGC\,1156
and NGC\,1569, the two nearest objects).
Solid dots refer to galaxies with corrected $H-K > 0.35$.
\label{fig:ciioikl} }
\end{figure}
%
the \oi\ line is expected to become more important relative
to \cii\ in warmer  and denser gas.
\cii\ and \oi\ line fluxes for the galaxies in our sample 
have been measured by ISO (Malhotra et al. \cite{m2001}), and in this section
we analyze those measurements in the context of our new photometry.

An important diagnostic is the ratio of \cii\ and the far-infrared
flux (FIR), since it measures essentially the efficiency of the
photoelectric heating of the gas by dust grain ejection.
This ratio \cii/FIR tends to decrease with warmer FIR colors
$F_{60}/F_{100}$ and increasing star-formation activity as indicated
by $FIR/B$ (Malhotra et al. \cite{m2001}).
Moreover, warmer gas, as signified by smaller \cii/\oi, correlates with warm
dust or larger $F_{60}/F_{100}$;
this last is the most significant correlation in the study by
Malhotra et al. (\cite{m2001}).
Since red $K-L$ should be related to hot dust and its temperature,
we might expect to find similar correlations with normalized
FIR line fluxes and ratios. 
However, this supposition is not borne out by the data
(see Fig. \ref{fig:ciioikl}):
we find no correlation between $K-L$ and
\cii/FIR, and only a weak anticorrelation between $K-L$ and \cii/\oi.
Again, $K-L$ appears to be measuring a hot-dust phase,
not closely connected to the properties of the cooler dust.

\section{PDR models  \label{pdr}}
\begin{figure}
\resizebox{\hsize}{!}{\rotatebox{0}{\includegraphics*[25,150][570,680]{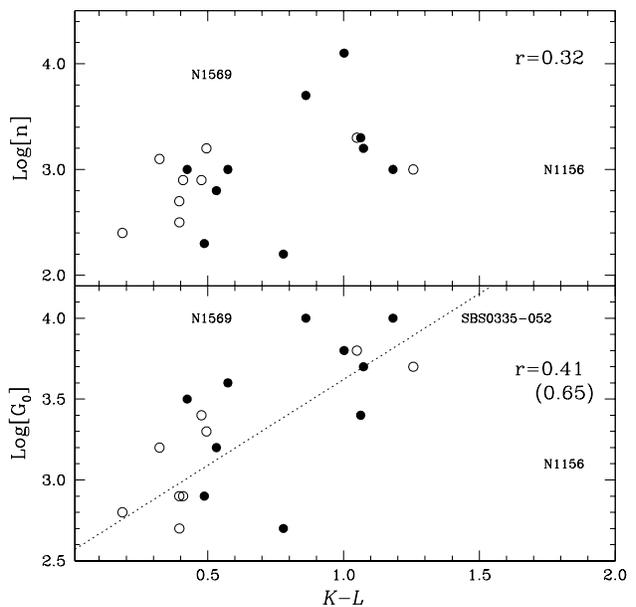}}}
\caption{Log\,$n$ (upper panel) and Log\,\g0\ (lower) plotted against
$K-L$.
The PDR parameters are taken from Malhotra et al. (\cite{m2001}).
The dotted line in the lower panel shows the best-fit regression;
the correlation coefficients are shown in each panel. Solid points
refer to galaxies with corrected $H-K > 0.35$. Estimate of 
correlation coefficients and regression include all objects
(the parentheses contain the coefficient without NGC\,1156 and NGC\,1569,
the two nearest galaxies).
\label{fig:ng0kl} }
\end{figure}
\begin{figure}
\resizebox{\hsize}{!}{\rotatebox{0}{\includegraphics*[25,150][570,680]{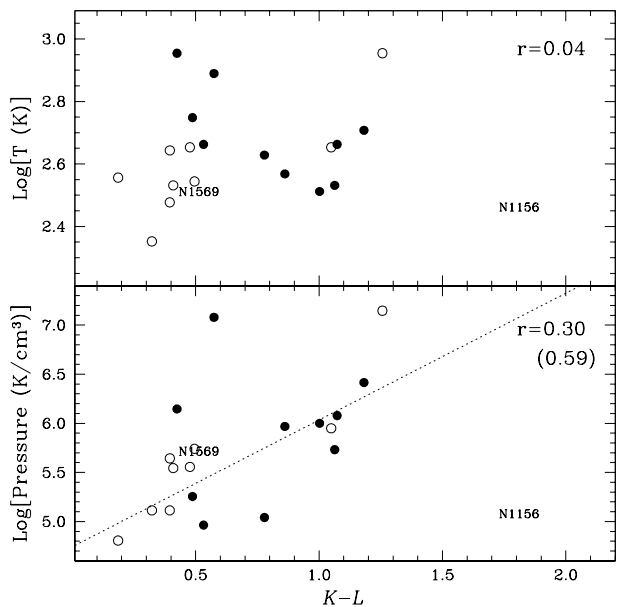}}}
\caption{Log\,$T$ (upper panel) and Log\,$P$ (lower) plotted against
$K-L$.
The PDR parameters are taken from Malhotra et al. (\cite{m2001}).
The dotted lines show the best-fit regression; the correlation coefficients are
reported in each panel. Solid points refer to galaxies with
corrected $H-K>0.35$. Estimates of regression line and correlation coefficients
include all galaxies (the parentheses contain the coefficient without the
two nearest objects, NGC\,1156 and NGC\,1569).
\label{fig:pTkl} }
\end{figure}
 
Photodissociation models for all the galaxies in our sample
have been reported in Malhotra et al. (\cite{m2001}), calculated from the
model grids of Kaufman et al. (\cite{kaufman}).
These models calculate the line emission in \cii\ and in \oi\
(including the line at 145\micron) and the dust continuum emission
for a plane slab of gas illuminated on one side by FUV radiation.
Gas heating in the models is dominated by photoelectrons ejected from
classical (large) grains and from VSGs according to the recipe
of Bakes \& Tielens (\cite{bakes}).
From these models, it is possible to infer values for the
neutral gas density $n$ and the FUV radiation field \g0, as well as for
the pressure $P$ and temperature $T$ of the gas.

Figure \ref{fig:ng0kl} shows gas density $n$ and FUV
flux \g0, as derived from PDR models, plotted against $K-L$. 
Without NGC\,1156 (one of the two nearest galaxies), $K-L$ is 
well correlated with \g0, at the 3.5\,$\sigma$ level.
This is a clear indication that red $K-L$ is connected to strong FUV
radiation fields.
Also shown in Fig. \ref{fig:ng0kl} is SBS\,0335-052, with
a \g0 $\,\sim\,$10000 times stronger than the local value
(Dale et al. \cite{dsbs}); this galaxy follows the same correlation.
 
We then examine whether the primary dependency of red $K-L$ color
is on neutral gas temperature $T$ or on its pressure $P$, given that the
$K-L$ color is independent, or roughly so, of neutral gas density $n$.
The relevant PDR parameters are plotted in  Fig. \ref{fig:pTkl} where
$T$ and $P$ as inferred from the PDR models are plotted versus $K-L$.
The thermal pressure $P$ of the PDR is expected to be approximately equal to
the thermal pressure of the adjacent \hii\ region (Malhotra et al. \cite{m2001}).
The correlation between $P$ and $K-L$ is weak but becomes significant
(2.8$\sigma$) once we exclude NGC 1156.
There is no apparent correlation between $T$ and $K-L$.
Taken together,
these results suggest that the PDR pressure (roughly equal to the \hii\
pressure) and the FUV radiation field \g0\ are the main factors which
govern the existence of hot dust in star-forming galaxies;
neutral gas density and temperature by themselves are less important influences.

\section{The nature of the 4\micron\ continuum \label{nature}}

We first discuss the nature of the 4\micron\ continuum, as inferred from
our observations.
Because of the filter cutoff,  \Lp\ misses the 3.3\micron\ aromatic
emission feature (e.g., Moorwood \& Salinari \cite{ms83}; 
Moorwood \cite{m86}), and thus should be a measure
of the AFE-free continuum.
As mentioned in Sec. \ref{colors}, the $K-L$ color is a sensitive diagnostic
of the fraction of hot dust contained in the observing aperture.
The 4\micron\ continuum of most of the galaxies we observed is consistent
with stellar photospheres together with moderate dust extinction.
As expected, the best cases for pure stellar $K-L$ colors are the ellipticals
which have median $K-L\,=\,0.21$.
Within the 14\arcsec\ observing aperture, we find no evidence for
circumstellar dust in the ellipticals, since their NIR colors are inconsistent
with dust emission at any temperature.

In roughly a third of our sample, the $K-L$ color ($\gtrsim\,1$)
signifies a flat or rising 4\micron\ continuum.
Even a small fraction (5\%) of hot (600\,K) dust can cause this inflection,
and is very likely the cause of the large variance in the 3--5\micron\
continua, as noted by Helou et al. (\cite{h2000}). 
Because of the characteristics of the \Lp\ filter,
we conclude that the red $K-L$ color must represent a continuum property,
rather than short-wavelength AFEs.
Such a continuum is typically accompanied by a lower $F_{6.75}/F_{15}$ ratio 
(see Fig. \ref{fig:gen}),
due either to AFE suppression or to an $F_{15}$ excess (or both).
Either way, we are observing a significant fraction of hot dust in the
central regions of these galaxies, with
physical conditions that favor red $K-L$, namely strong \g0\ and
high pressure.

\subsection{Physical conditions in star forming regions 
with hot dust \label{physical}}

Malhotra et al. (\cite{m2001}) describe a physical picture in which the infrared
line emission arises in a PDR which surrounds an expanding \hii\ region.
They suggest that the ISO diagnostics are probing the typical distance
from an OB star or cluster to the PDR gas,
rather than the global rate of star formation.
Our new 4\micron\ data are consistent with this picture.
Red $K-L$ tends to be associated with an intense UV radiation field \g0.
The trend shown in Fig. \ref{fig:ng0kl} suggests that rising 4\micron\
continua are observed when UV radiation field \g0 $\,\gtrsim\,10^{3.5}$.
It is easy to show that such an intense \g0\ must be found within a few tens of
pc from the ionizing star cluster.
Assuming the FUV luminosity of a single star to be
$10^{39}$\,erg s$^{-1}$ (!),
and with a star cluster consisting of 100 OB stars, 
the typical distance for \g0\,=\,$10^{3.5}$ is 10\,pc.
In the same conditions, at a distance of 100\,pc we would need 10000 massive
stars, more than all but those in the most luminous Super Star Clusters
(SSCs: e.g., Calzetti et al. \cite{calzetti2}, Turner et al. \cite{turner} ).
With the lower FUV luminosity from lower-mass stars, the distance to maintain
the intense \g0\ would be even smaller. 
Hence, it is likely that red $K-L$ is probing a region within a few tens
of pcs from star clusters containing a few hundred OB stars.
Such sizes are more than a hundred times smaller than the projected size
of our observing aperture at the median distance of the sample
(see Sect. \ref{obs}).

Hot dust as measured by high-resolution 10--12\micron\ images of
infrared-luminous starbursts also tends to be compact, and compact nuclear
starbursts generally dominate their starburst activity (Soifer et al. \cite{soifer}).
Typical MIR sizes in the Soifer et al. sample
range from $<\,$125\,pc to $\sim\,$1\,kpc; however these may be partly
overestimated since the spatial resolution for the nearest galaxy in
their study (NGC\,3690) is 210\,pc\,arcsec$^{-1}$ which procured mostly
upper limits ($<\,$125\,pc) for the size of the emitting regions.

The hot dust that gives rise to red $K-L$ must be heated by massive stars,
and according to the PDR models, tends to be found in high-pressure
environments and, from the above discussion, 
in close proximity of the \hii\ region.
This intense environment suppresses AFE emission, 
probably because of the strong UV radiation field, as
proposed by earlier work (e.g., Normand et al. \cite{normand}). 
In the framework of an expanding \hii\ region surrounded by a PDR,
such conditions are likely to obtain in young \hii\ regions, still embedded
within their natal molecular cloud, which have not yet had time to break
through the cloud surface. 
At first glance then, $K-L$ appears to be a measure of age:
hot dust is found preferentially in young \hii\ regions. 

Nevertheless, the attribution of the rising 4\micron\ continuum to youth
has some problems.
Some very young systems have ``normal'' photospheric$+$extinction $K-L$ colors:
the two brightest star clusters in NGC\,1569 contain Wolf-Rayet (W-R) stars and their
age has been estimated at $<$\,5\,Myr (Origlia et al. \cite{origlia}).
We measure for NGC\,1569 $K-L\,=\,0.7$, consistent with 
stellar photospheres $+$ extinction, not with hot dust.
Haro\,2 (not in this sample) also contains W-R stars
(Vacca \& Conti \cite{vacca}), and as such must be younger than 5\,Myr,
but $K-L=0.5$ (Thuan \cite{thuan83}).
Thus, young systems are not necessarily associated with 
red $K-L$ and hot dust.
\begin{figure}
\resizebox{\hsize}{!}{\rotatebox{0}{\includegraphics*[25,150][570,680]{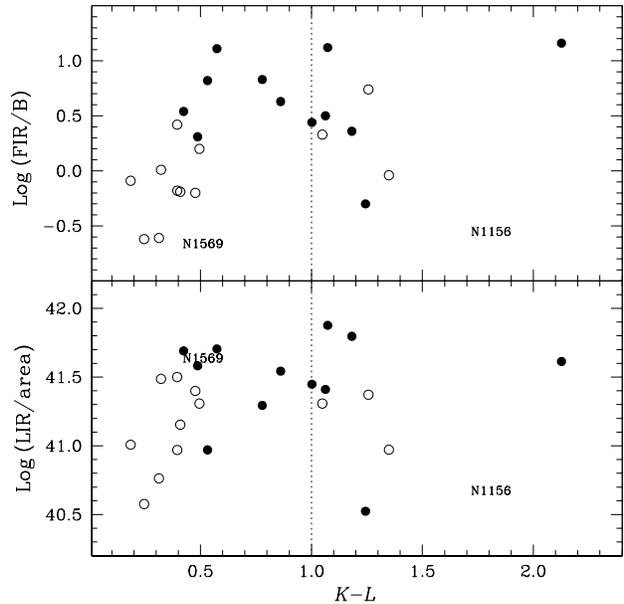}}}
\caption{
Top panel: Log of the far-infrared-to-blue flux ratio $FIR/B$ vs. $K-L$. 
$FIR/B$ was taken from Dale et al. (\cite{d2000}), with $FIR$
calculated according to the canonical dependence on
$F_{60}$ and $F_{100}$ (Lonsdale Persson \& Helou \cite{persson87}).
Bottom panel: Log\,$L_{FIR}/{\rm area}$ versus $K-L$.
$L_{FIR}$ is the far-infrared luminosity and
the area is the one of the
ellipse with axes of the size reported in NED;
$L_{FIR}/{\rm area}$ is in erg\,s$^{-1}$\,kpc$^{-2}$.
Galaxies with corrected $H-K\,\leq\,0.35$ are shown as open circles,
and $H-K\,>\,0.35$ as filled ones.
The vertical dotted line separates the ``active'' and ``passive'' regimes
($K-L\,\gtrsim\,1$).
\label{fig:firkl} }
\end{figure}

It also might be that $K-L$ is a diluted version of the flux ratio $FIR/B$, since 
variations in $FIR/B$, or star-formation activity,
might be translated into variations of the hot-dust 
versus stellar fraction in our observing aperture.
We investigate this in the top panel of Fig. \ref{fig:firkl}, where we have plotted 
Log($FIR/B$) vs. $K-L$.
There is no correlation between these two quantities, although
the galaxy with the highest value of $FIR/B$ is associated with 
the reddest $K-L$ (IRAS\,23365+3604).
It is thus difficult to interpret $K-L$ as an indicator of
generic star formation, since the hot dust traced by $K-L$ 
appears to be independent of the cooler grains traced by FIR.
\subsection{``Active'' and ``passive'' star formation \label{active}}
We therefore propose an alternative explanation, namely two ``extreme'' distinct
modes of star formation. 
Recent models have shown that dust heating and molecular hydrogen production are
more efficient in dense ($\gtrsim$\,500\,cm$^{-3}$), compact 
($\ltrsim$\,100\,pc) environments (Hirashita et al. \cite{hirashita2002}).
Moreover, in low-metallicity BCDs, the size of the brightest star-forming complex and
its ionized gas density are very well correlated, with the densest regions being
also the most compact (Hunt et al. \cite{hunt2002}).
In this same BCDs sample, size and density are independent of age, since ages derived
from recombination line equivalent widths are all $\ltrsim$\,5\,Myr. 
The correlations in the KP sample studied here seem to suggest a similar
``dichotomy'' since a strong UV radiation field, compact size, and pressure
are associated with hot dust.
The most compact dense regions in the BCD study by Hunt et al. (\cite{hunt2002})
were interpreted by them as indicative of ``active'' star formation,
characterized by an intense physical environment 
harboring several hundreds/thousands of massive stars, and 
capable of efficiently heating dust. 
At a comparable star-formation rate,
the less dense, less compact \hii\ regions, termed ``passive''
by Hunt et al. (\cite{hunt2002}), are not able to heat dust
efficiently because of the more diluted radiation field (Hirashita et al. \cite{hirashita2002}). 
We would argue that a rising 4\micron\ continuum
is yet another signature of ``active'' star formation.
A falling 4\micron\ continuum would be a sign of ``passive'' star formation,
since the physical conditions are not able to produce a measurable fraction
($\gtrsim\,5$\%) of hot dust; these more diffuse regions tend to have
lower pressures, a less intense radiation field, and prominent AFEs in the 
MIR spectrum.

It might be argued that the two extremes are simply 
a measure of the star-formation rate (SFR).
We have attempted to test this hypothesis by plotting the far-infrared
luminosity per unit area $L_{FIR}/{\rm area}$ against
$K-L$, since $L_{FIR}$ is frequently used to measure SFRs (e.g.,
Thronson \& Telesco \cite{thronson}). 
The bottom panel of Fig. \ref{fig:firkl} shows that
the two quantities are not significantly correlated,
nor is $K-L$ correlated with $L_{FIR}$, that is without
normalizing by the area (not shown).
We already said that, although aperture effects may be important
because of the large IRAS apertures,
the top panel of the same figure shows no correlation between
$K-L$ and the $L_{FIR}$ {\it normalized} to the $B$ band,
nor with $F_{60}/F_{100}$ (Sect. \ref{isoflux}), yet
another star formation indicator of common use.
It appears, therefore, that the difference between the two extremes is
of qualitative rather than quantitative origin.

It should be emphasized that ``active'' and ``passive'' star formation 
as we are defining them are {\it local} concepts:
in a single galaxy, different regions may be characterized by either mode.
In our Galaxy AFEs are notably diminished in \hii\ regions (Cesarsky et al. \cite{cesarsky98}),
and the MIR continuum tends to be stronger in denser structures (Abergel et al. \cite{abergel}).
M\,82 is an another example, since where the infrared continuum peaks,
there are few if any AFEs, but at the \hii\ region-molecular cloud
interface, the AFEs dominate (Normand et al. \cite{normand}).
On galactic scales, there is also evidence that AFEs are diminished as
the intensity of star formation increases
(F\"orster Schreiber et al. \cite{natascha}).
With our ground-based data, 
we are investigating the properties of the dominant mode of star
formation in the observing aperture, weighted by the brightness
of the components.
At longer wavelengths, higher spatial resolution is necessary to separate
the two modes because in large apertures (e.g., IRAS, ISO),
the cooler ``standard'' ISM virtually overwhelms any hot dust emission.

We further note that the bulk of star formation associated with starbursts 
is not necessarily of the ``active'' type.
AFEs are always detected in galaxy nuclei with \hii-region 
spectra (Roche et al. \cite{roche}); indeed
these features dominate the integrated spectra of prototypical starbursts
such as M\,82 and NGC\,253 (Sturm et al. \cite{sturm}).
They are also used to distinguish star formation from nuclear activity
as the mechanism powering ultra-luminous infrared galaxies (ULIRGs,
Genzel et al. \cite{genzel}).
This apparent ``paradox'' can be understood through the association of AFEs
with the cool dust phase of the PDR (Haas et al. \cite{haas}).
The hot dust we measure with $K-L$ is heated by the intense \g0\ close
to the \hii\ region, and without high-resolution ($\ltrsim\,100$\,pc)
measurements, the copious emission from the diffuse cooler dust
and AFEs in the PDRs dominates the observed flux.

We have found evidence for hot dust in roughly 1/3 of the KP galaxies
observed.  It is not clear, however, whether the properties of these
``active'' star-forming galaxies are ``extreme'' compared to other
examples of nuclear starbursts.
An upper limit to the bolometric surface brightness of these objects can be 
deduced from the estimates given in Sec. \ref{physical}:  if we have a nuclear star
cluster containing 100--1000 massive stars concentrated in a region of 10--100\,pc
in diameter, the bolometric surface brightness would be roughly
$\ltrsim\,10^{11}-10^{12}\,L_\odot$\,kpc$^{-2}$.  This coincides
approximately with the maximum mean surface brightness in the
sample of low- and high-redshift starbursts
of (Meurer et al. \cite{meurer}).  
However, our estimate refers to the {\it nuclear star clusters},
not to the mean surface brightness over the observing aperture as in
Meurer et al.
As star clusters go,
those we have detected with $K-L$ are not extraordinary,
since they are several times less bright than the resolved
UV clusters described by Meurer et al. (\cite{m95}).

What we have called the ``active'' mode is characterized by 
the formation of compact star-forming complexes 
within a high-pressure environment in an intense \g0.
High pressure is also implicated in the strong galactic winds 
common in starbursts (Heckman et al. \cite{heckman}).
Such conditions also favor the formation of dusty SSCs
(Bekki \& Couch \cite{bekki}),
which are frequently found in merging galaxies
(e.g., the Antennae: Whitmore et al. \cite{whitmore};
NGC\,1741: Johnson et al. \cite{johnson}),
ULIRGs (Scoville et al. \cite{scoville}, Shioya et al. \cite{shioya}),
and nuclear starbursts
(e.g., M\,82: Gallagher \& Smith \cite{gallagher}; 
NGC\,253: Keto et al. \cite{keto}; NGC\,4214: Leitherer et al. \cite{leitherer}).
Although somewhat uncertain, the nuclear star clusters inferred from our
data have properties similar to SSCs, which could be yet
another signature of ``active'' star formation.

\subsection{Speculations for high-redshift studies}
The physical environment of the ``active'' mode, especially high
pressure, results naturally from interactions or mergers (Bekki \& Couch \cite{bekki}).
Interestingly, interactions and mergers also produce red $K-L$ (Joseph et al. \cite{joseph}). 
According to the Cold Dark Matter hierarchical clustering scenario of
galaxy formation,
the frequency and intensity of interactions are expected
to increase with redshift,
and semi-analytic recipes for galaxy formation based on 
``collisional starbursts'' are successful at reproducing SFRs, colors, morphology, 
and mass trends with redshift (Somerville et al. \cite{somerville}).
Presently, the bulk of star formation in galaxies, including
starburst galaxies and ULIRGs, takes place in the ``passive'' regime.
But, due to the higher probability of strong tidal stresses and mergers,
this situation may have been reversed in the past with the ``active'' 
regime being the preferred way to form stars at high redshift.

In this context, it is puzzling that we maintain that ULIRGs host 
passive star formation, although most if not all of them are mergers.
The most plausible explanation of this paradox involves spatial resolution
and the relative dominance of the ``passive'' ISM in ULIRGs (see Sect.
\ref{active}).
Given the large contribution from their cool ISM, the emission in ULIRGs 
integrated over large spatial scales appears passive. 
Even though ``active'' star formation must be occurring,
it can be revealed only on small spatial scales, or
by its hot dust.
A case in point is the ULIRG in our sample, IRAS\,23365$+$3604.
This galaxy hosts several young ($\leq$\,10\,Myr) star clusters
(Surace et al. \cite{surace}), but is classified optically as a LINER
(Veilleux et al. \cite{veilleux}).
The LINER classification is probably not caused by an AGN, but rather by
shocks generated in galactic superwinds (Lutz et al. \cite{lutz}).
We classify this object as ``active'' because of its extremely red $K-L$\footnote{The red
$K-L$ cannot easily be attributed to nuclear activity, since LINERs tend to
have NIR colors dominated by normal stellar populations (Lawrence et al. 
\cite{lawrence}).},
but its ISO spectrum contains prominent AFEs (Tran et al. \cite{tran}).
The young star clusters probably heat the dust responsible for the
red $K-L$, while the ISO aperture encompasses the entire
galactic disk (see Surace et al. \cite{surace}).
Hence, IRAS\,23365$+$3604 appears both active and passive: active because of
its red $K-L$, and passive because of the overwhelming disk ISM. 

Including the effects of active star formation in models of galaxy evolution
(e.g., Granato et al. \cite{granato}) would not be straightforward.
Nevertheless, when star formation occurs on spatial scales $\ltrsim$\,100\,pc 
(see Sect. \ref{physical}), the intense radiation field, high pressure, and diminished
AFEs should be taken into account.
Indeed, suppressed AFEs coupled with the hot dust revealed by our photometry 
substantially alter the infrared spectrum of star-forming galaxies.
This may be especially important in hierarchical merger models because of the supposed link
between ``active'' star formation and interactions.

\section{Conclusions \label{conclusions}}
With our $JHK$\Lp\ photometry, we have analyzed the 4\micron\ continuum and its
relation with the MIR spectrum and FIR emission lines.
We find the following:
\begin{enumerate}
\item 
The majority of the 26 KP galaxies have a falling 4\micron\ continuum,
$K-L\,<\,1.0$, consistent with stellar photospheres and moderate dust extinction.
10 of them have a flat or rising 4\micron\ continuum, $K-L\,\gtrsim\,1.0$, 
consistent with a measurable fraction of 600--1000\,K hot dust.
\item 
$K-L$ is anticorrelated with ISO ratios $F_{6.75}/F_{15}$ and IRAS ratio 
$F_{12}/F_{25}$, but only weakly with \cii/\oi, 
and not at all with \cii/FIR/ or IRAS ratio $F_{60}/F_{100}$.
\item
PDR models for these galaxies show that the hot dust measured by red $K-L$
is associated with high pressures and intense far-ultraviolet radiation fields
in compact ($\ltrsim\,$100\,pc) regions.
\item
These results taken together suggest that star formation in these galaxies
occurs in two ``extreme forms'':
\begin{enumerate}
\item
a relatively rare ``active'' mode characterized by hot dust,
suppressed AFEs, high pressure, intense ultraviolet radiation field,
and compact size;
\item
a more common ``passive'' mode characterized by photospheric $K-L$ colors,
with moderate extinction, and less extreme physical conditions.
\end{enumerate}
\item
The physical conditions we infer for the star-forming regions containing
hot dust are similar to those created by interactions and mergers.
We speculate that such intense episodes may have been more 
common in the past,
so that the ``active'' regime could dominate star formation at high redshift.
\end{enumerate}
\begin{acknowledgements}
The TIRGO staff was instrumental in the success of these observations.
We are grateful to Giovanni Moriondo who valiantly attempted to observe
for this project, and to the TIRGO Time Allocation
Committee for generous time allocations.
We gladly acknowledge enlightening discussions with Marc Sauvage, {\bf and
would like to thank an anonymous referee for several insightful comments}.
\end{acknowledgements}

\end{document}